\documentclass[a4paper,11pt]{article}
\usepackage{pos}
\usepackage{bibspacing}

\newcommand{\tcite}[1]{~\cite{#1}}
\newcommand{\tref}[1]{~\ref{#1}}
\newcommand{\eref}[1]{~\eqref{#1}}

\title{Toward twist-2 $T$-odd transverse-momentum-dependent gluon distributions: the $f$-type linearity function
}
\ShortTitle{Toward twist-2 $T$-odd TMD gluon distributions: the linearity function}

\author[a,b]{Alessandro Bacchetta}
\author*[c,d,e]{Francesco Giovanni Celiberto}
\author[b]{Marco Radici}

\affiliation[a]{Dipartimento di Fisica, Universit\`a di Pavia, via Bassi 6, I-27100 Pavia}
\affiliation[b]{INFN Sezione di Pavia, via Bassi 6, I-27100 Pavia, Italy}
\affiliation[c]{European Centre for Theoretical Studies in Nuclear Physics and Related Areas (ECT*), I-38123 Villazzano, Trento, Italy}
\affiliation[d]{Fondazione Bruno Kessler (FBK), I-38123 Povo, Trento, Italy}
\affiliation[e]{INFN-TIFPA Trento Institute of Fundamental Physics and Applications, I-38123 Povo, Trento, Italy}

\emailAdd{alessandro.bacchetta@unipv.it}
\emailAdd{fceliberto@ectstar.eu}
\emailAdd{marco.radici@pv.infn.it}

\abstract{
We present preliminary results on the leading-twist transverse-momentum-dependent (TMD) gluon linearity function, which is directly connected to nucleon transverse-spin asymmetries originating from the density of linearly-polarized gluons.
The function is calculated in a spectator-model framework for the parent proton and for the $f$-type gauge-link structure.
Our work represents a further step toward the definition of a complete set of gluon TMD distributions at twist-2 that can be used to access the gluon dynamics inside nucleons and nuclei at new-generation colliders.
}

\FullConference{%
  *** Particles and Nuclei International Conference - PANIC2021 ***\\
  *** 5 - 10 September, 2021 ***\\
  *** Online ***
}

\begin{document}
\maketitle

\section{Introductory remarks}

\vspace{-0.00cm}

The study of the inner structure of nucleons in terms of the distribution of their constituents has always represented a frontier research field in particle physics.
The well-established \emph{collinear} factorization has played an important role in the description of high-energy hadronic and lepto-hadronic collisions in terms of \emph{one-dimensional} parton distribution functions (PDFs).
There are fundamental questions, however, that are still open and whose answers cannot be made from a purely collinear perspective.
To shed light on the origin of proton spin and mass as well as on azimuthal asymmetries determined by the interplay between nucleon and parton polarizations, a \emph{three-dimensional} vision is required. 
The \emph{transverse-momentum-dependent} (TMD) factorization represents the most powerful tool to afford such a \emph{tomographic} description.
At variance with the quark-TMD sector, where significant results have been collected both on the formal and phenomenological sides, gluon TMDs are still a largely unexplored territory.
The first classification of (un)polarized TMD gluon densities was done in~\cite{Mulders:2000sh} and then extended in~\cite{Meissner:2007rx,Lorce:2013pza,Boer:2016xqr}. First phenomenological studies were conducted in~\cite{Lu:2016vqu,Lansberg:2017dzg,Gutierrez-Reyes:2019rug,Scarpa:2019fol,Adolph:2017pgv,DAlesio:2017rzj,DAlesio:2018rnv,DAlesio:2019qpk}.
A remarkable feature that distinguishes TMD densities from their collinear counterparts is the sensitivity to the gauge link, which leads to the well-known TMD process dependence~\cite{Brodsky:2002cx,Collins:2002kn,Ji:2002aa}.
While quark TMDs depend on processes via the $[+]$ and $[-]$ staple links respectively depicting the direction of future- and past-pointing Wilson lines, gluon TMDs depend on combinations of those gauge links, thus bringing to a more diversified \emph{modified universality}.
Two major gluon gauge links appear: the $f\text{-type}$ (or Weisz\"acker--Williams) and the $d\text{-type}$ (or dipole) ones.
The $f$-type $T$-odd gluon-TMD correlator is characterized by the $f_{abc}$ QCD color structure, whereas the symmetric $d_{abc}$ structure emerges in the $d$-type $T$-odd one. It turns out that $f$-type TMDs depend on $[\pm,\pm]$ gauge links, while $d$-type TMDs depend on $[\pm,\mp]$ ones.
Box-loop gauge links are probed via reactions where multiple color exchanges connect both initial and final states~\cite{Bomhof:2006dp}, thus generating factorization-breaking effects~\cite{Rogers:2013zha}.
The connection between TMD and \emph{high-energy} factorization was investigated in recent studies on the BFKL unintegrated gluon density~\cite{Bolognino:2018rhb,Bolognino:2018mlw,Bolognino:2019bko,Bolognino:2019pba,Celiberto:2019slj,Celiberto:2018muu,Celiberto:2020wpk,Bolognino:2019yls,Bolognino:2021niq,Celiberto:2020tmb,Celiberto:2020rxb,Bolognino:2021mrc,Celiberto:2021dzy,Celiberto:2021fdp,Celiberto:2022dyf,Nefedov:2021vvy,Hentschinski:2021lsh}.
Detailed studies on spectator-model quark TMDs in the proton were done in\tcite{Bacchetta:2008af,Bacchetta:2010si}. 
A common framework was recently built\tcite{Bacchetta:2020vty} (see also\tcite{Bacchetta:2021oht,Celiberto:2021zww}) for all the $T$-even gluon TMDs at twist-2. These functions were calculated in an enhanced spectator model for the parent proton and they encode effective high-energy resummation effects.
In\tcite{Bacchetta:2021lvw} we presented a first extension of our framework to $T$-odd gluon TMDs by means of the $f$-type Sivers function, which carries information on the distribution of unpolarized gluons inside transversely polarized protons.
In this work we present a preliminary analysis of another $T$-odd gluon TMD, the $f$-type linearity function, which can explain observed single-spin asymmetries originating from the density of linearly-polarized gluons inside transversely polarized protons.

\vspace{-0.00cm}

\section{The $f$-type linearity function in a spectator model}

\vspace{-0.15cm}

\begin{figure}[t]
\centering
\includegraphics[width=0.43\textwidth]{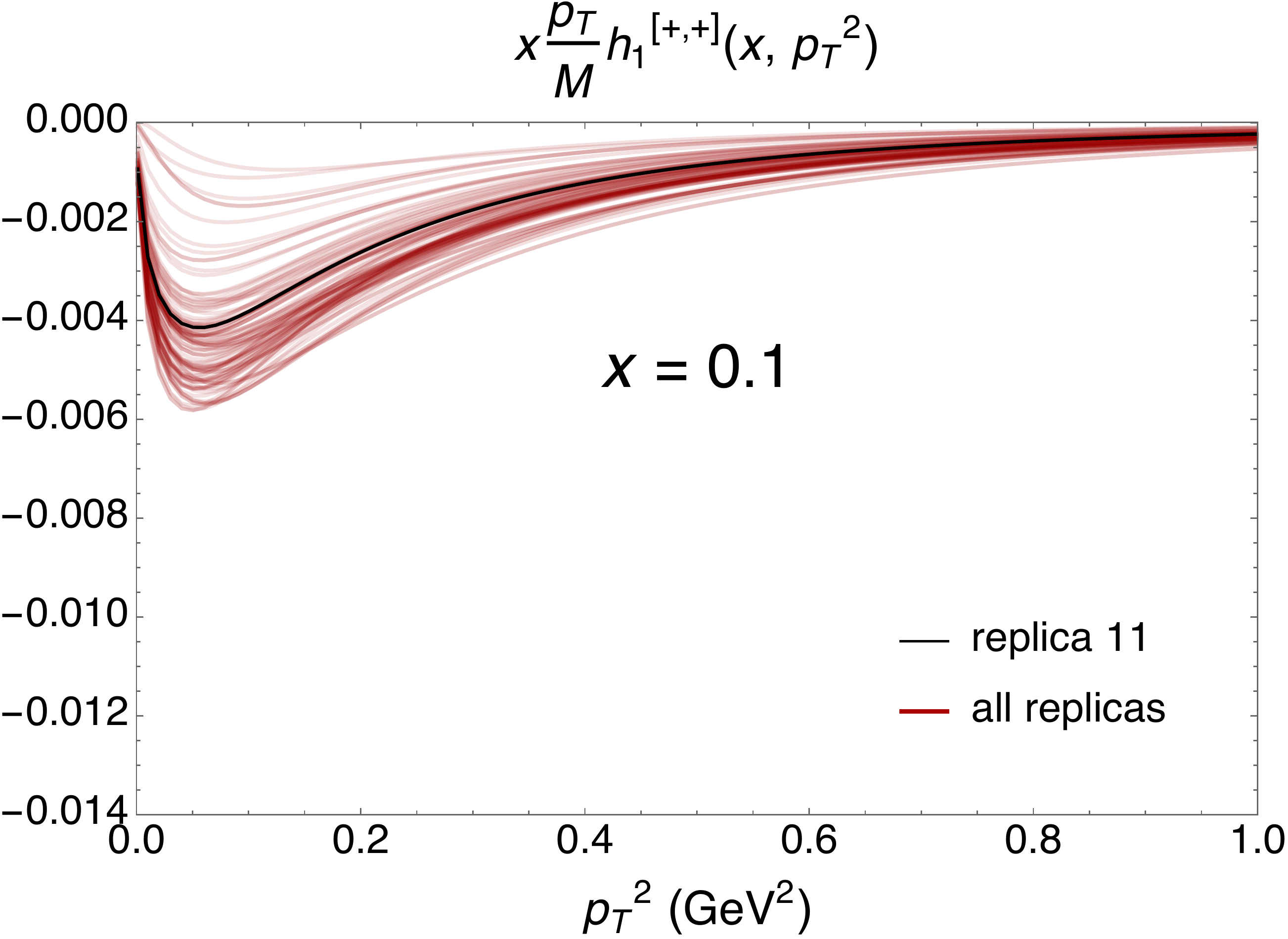} \hspace{0.5cm}
\includegraphics[width=0.43\textwidth]{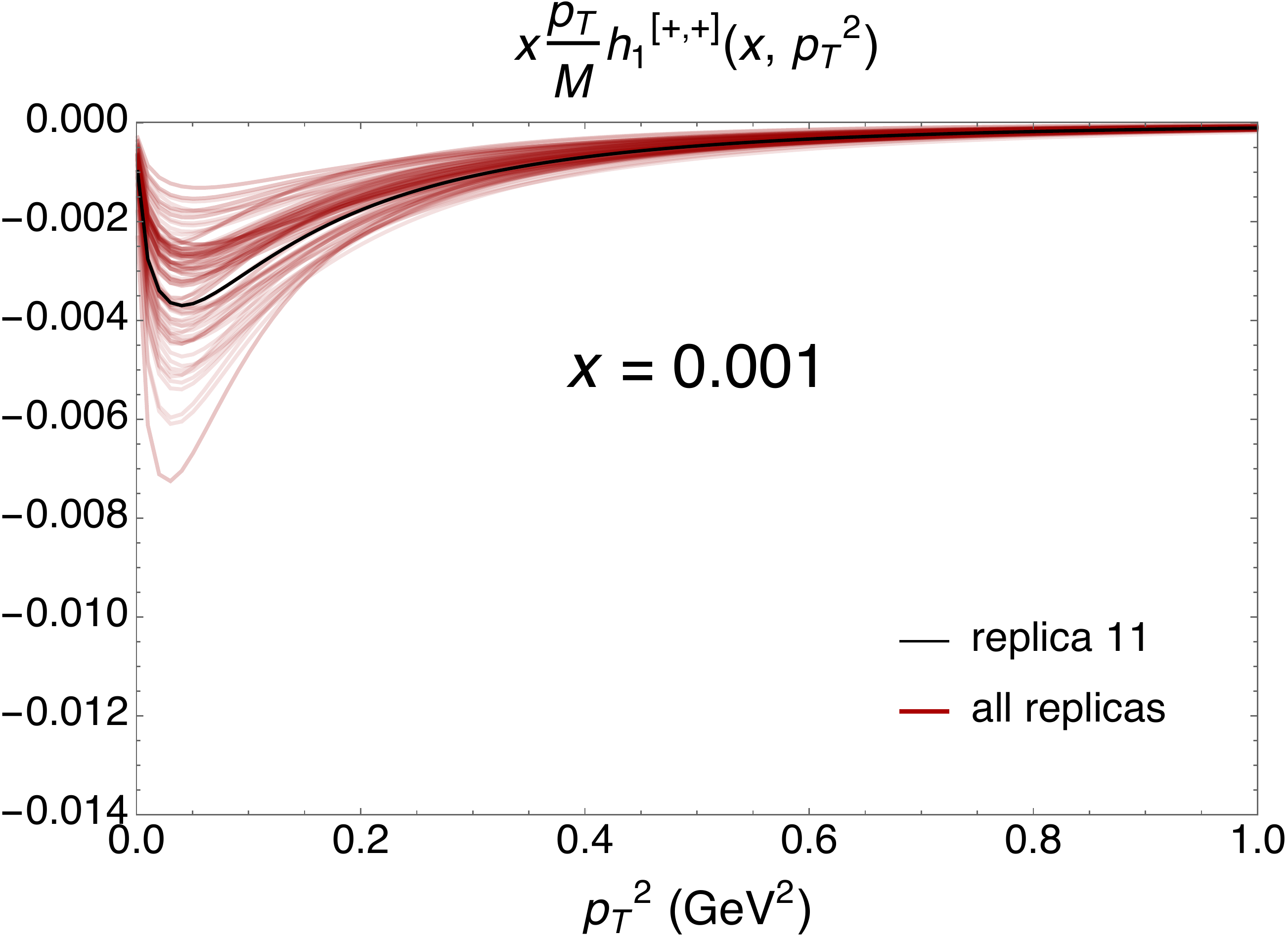}
\caption{Transverse-momentum dependence of the $[+,+]$ linearity function for $x=10^{-1}$ (left) and $x=10^{-3}$ (right), and at the initial scale $Q_0 = 1.64$ GeV. Black curve stands for the most representative replica \#11.}
\label{fig:linearity_f}
\end{figure}

The spectator-model framework relies on the assumption that an incoming nucleon, having mass $M$ and four-momentum $P$, emits a parton with four-momentum $p$, transverse momentum $\boldsymbol{p}_T$ and longitudinal fraction $x$, and what remains is considered as an effective on-shell spectator particle having mass $M_X$ and spin-1/2.
We model the nucleon-gluon-spectator vertex in the following way
\begin{equation}
 \label{eq:form_factor}
 {Y}^\mu = \left( g_1(p^2)\gamma^\mu + g_2(p^2) \frac{i}{2M} \sigma^{\mu\nu}p_\nu \right) \; .
\end{equation}
In Eq.\eref{eq:form_factor} $g_1$ and $g_2$ are dipolar functions of $\boldsymbol{p}_T^2$. Employing dipolar form factors is quite useful, since it permits us to quench gluon-propagator divergences, suppress large-$\boldsymbol{p}_T$ effects that are beyond the reach of a genuine TMD description, and cancel logarithmic singularities arising from $\boldsymbol{p}_T$-integrated densities.
All the leading-twist $T$-even gluon TMDs in the proton were calculated in\tcite{Bacchetta:2020vty} by defining an enhanced version of the tree-level spectator model. 
In particular, the $M_X$ mass of the colored spin-1/2 spectator was integrated over a continuous range weighed by a flexible spectral function suited to catch both small- and moderate-$x$ effects (see Eqs.~(16) and~(17) of~\cite{Bacchetta:2020vty}).
Parameters embodied by the spectral mass and the pure spectator-model correlator were fixed by making a simultaneous fit of our unpolarized and helicity TMD distributions, $f_1^g$ and $g_1^g$, to the corresponding collinear PDF functions from {\tt NNPDF}\tcite{Ball:2017otu,Nocera:2014gqa} at the initial scale $Q_0 = 1.64$ GeV. Statistical uncertainties are described via the bootstrap method.
The spectator-model $T$-even functions are process-independent, since the tree-level approximation for the gluon correlator does not account for the gauge link.
The interference term between two scattering amplitudes with different imaginary parts is needed to generate any $T$-odd structure. The most straightforward option is to go beyond the tree-level gluon correlator by including the interference with the one-gluon exchange process (in \emph{eikonal} approximation). The latter represents the first-order approximation of the gauge link operator. The main consequence of this strategy is that the calculated $T$-odd TMDs is gauge-link sensitive, and thus process dependent. The two linearity functions corresponding to the $f$-type gauge link are obtained by suitably projecting the transverse part of the corresponding gluon correlator, and they are equal up to a minus sign
\begin{equation}
 \label{eq:linearity_f}
 h_1^{g \, [+,+]}(x, \boldsymbol{p}_T^2) \equiv - h_1^{g \, [-,-]}(x, \boldsymbol{p}_T^2) \; .
\end{equation}
In our preliminary analysis we use a simplified expression of the nucleon-gluon-spectator vertex, where the $g_2$ form factor in Eq.\eref{eq:form_factor} is set to zero.
For the sake of consistency, we fitted again model parameters to {\tt NNPDF} results by using the simplified formula for the vertex.
Fig.\tref{fig:linearity_f} shows the $\boldsymbol{p}_T^2$-shape of the $[+,+]$ linearity function at $x = 10^{-1}$ and at $x = 10^{-3}$, and at the initial scale $Q_0 = 1.64$ GeV.
The absolute value of our density is clearly a non-Gaussian function of $\boldsymbol{p}_T^2$, with a large flattening tail at large $\boldsymbol{p}_T^2$-values and a small nonzero value when $\boldsymbol{p}_T^2$ goes to zero.
We remark that this behavior could radically change  when the full-vertex calculation will become available.

\vspace{-0.12cm}

\section{Closing statements}

\vspace{-0.15cm}

We have extended our spectator-model framework by including a preliminary version of the $f$-type gluon linearity TMD function. Once the calculation of all the $T$-odd gluon TMDs will be completed, we will include standard TMD evolution and perform phenomenological analyses on the relevant spin asymmetries that can be studied at new-generation colliding facilities, as the \emph{Electron-Ion Collider}~(EIC)~\cite{AbdulKhalek:2021gbh}, the \emph{High-Luminosity Large Hadron Collider} (HL-LHC)~\cite{Chapon:2020heu}, NICA-SPD~\cite{Arbuzov:2020cqg}, and the \emph{Forward Physics Facility} (FPF)~\cite{Anchordoqui:2021ghd}.

\vspace{-0.00cm}

\bibliographystyle{bibstyle}
\bibliography{references}

\end{document}